\pdfoutput=1
\documentclass[12pt]{iopart}
\usepackage{iopams}
\usepackage{graphicx}
\usepackage{gensymb}
\usepackage{enumerate}
\usepackage{stmaryrd}
\usepackage{amssymb}
\usepackage{wasysym}
\usepackage{hyperref}

\begin{document}

\title[Susceptibility divergence, phase transition and multistability of a turbulent flow]{Susceptibility divergence, phase transition and multistability of a highly turbulent closed flow}

\author{P-P~Cortet$^{1,2}$, E~Herbert$^1$, A~Chiffaudel$^1$, F~Daviaud$^1$, B~Dubrulle$^1$ and V~Padilla$^1$}

\address{$^1$ CEA Saclay, IRAMIS, SPEC, CNRS URA 2464, Groupe Instabilit\'{e}s \& Turbulence, Orme des Merisiers, 91191 Gif-sur-Yvette, France}
\address{$^2$ Laboratoire FAST, CNRS, Univ Paris-Sud, UPMC Univ Paris 06, B\^at. 502, Campus universitaire, 91405 Orsay, France}

\ead{arnaud.chiffaudel@cea.fr}

\begin{abstract}
Using time-series of stereoscopic particle image velocimetry data,
we study the response of a turbulent von K\'{a}rm\'{a}n swirling
flow to a continuous breaking of its forcing symmetry. Experiments
are carried over a wide Reynolds number range, from laminar regime
at $Re = 10^{2}$ to highly turbulent regime near $Re = 10^{6}$. We
show that the flow symmetry can be quantitatively characterized by
two scalars, the global angular momentum $I$ and the mixing layer
altitude $z_s$, which are shown to be statistically equivalent.
Furthermore, we report that the flow response to small forcing
dissymetry is linear, with a slope depending on the Reynolds
number: this response coefficient increases non monotonically from
small to large Reynolds number and presents a divergence at a
critical Reynolds number $Re_c = 40\,000 \pm 5\,000$. This
divergence coincides with a change in the statistical properties
of the instantaneous flow symmetry $I(t)$: its pdf changes from
Gaussian to non-Gaussian with multiple maxima, revealing
metastable non-symmetrical states. For symmetric forcing, a peak
of fluctuations of $I(t)$ is also observed at $Re_c$: these
fluctuations correspond to time-intermittencies between metastable
states of the flow which, contrary to the very-long-time-averaged
mean flow, spontaneously and dynamically break the system
symmetry. We show that these observations can be interpreted in
terms of divergence of the susceptibility to symmetry breaking,
revealing the existence of a phase transition. An analogy with the
ferromagnetic-paramagnetic transition in solid-state physics is
presented and discussed.
\end{abstract}

\noindent \textbf{Keywords:} Turbulence, Metastable states,
Critical exponents and amplitudes (Experiments), Classical phase
transitions (Experiments)

\maketitle

\section{Introduction}

Symmetry breaking is an essential ingredient of classical phase
transitions \cite{landau}. Remarkably, it also governs the
transition to turbulence, that usually proceeds, as the Reynolds
number $Re$ increases, through a sequence of bifurcations breaking
successively the various symmetries allowed by the Navier-Stokes
equations coupled to the boundary conditions \cite{manneville}.
Finally, at large Reynolds number, when the fully developed
turbulent regime is reached, all the broken symmetries are
restored in a statistical sense \cite{frisch}.

Beside this well-established scenario, some experiments in closed
flows have raised intriguing features proceeding during transition
to turbulence
\cite{chilla2004,mujica2006,gibert2009,lohse2010,tabeling96,tabeling02,ravelet2004,ravelet2008}.
For instance, Tabeling \textit{et al.} have evidenced a local peak
in the flatness of velocity derivative in a von K\'{a}rm\'{a}n
flow of helium around $Re=2 \times 10^5$ \cite{tabeling96} that
was claimed to exhibit some characteristics of a second-order
phase transition \cite{tabeling02} and suggested to be associated
with the breakdown of small-scale vortical structures. However,
since then, no numerics or experiments ever confirmed this
scenario, leaving this flatness peak as an unsolved problem. More
recently, a transition has been evidenced in another von
K\'{a}rm\'{a}n flow around $Re=10^4$
\cite{ravelet2004,ravelet2008} which consists in a global
bifurcation of the mean flow from the basic symmetric flow
topology towards two other flow topologies which spontaneously
break the symmetry of the driving apparatus. This transition is
associated with an hysteresis and a divergence of transition
times, two features classically observed in phase transitions.
Even more recently, in a von K\'{a}rm\'{a}n flow very similar to
the flow considered here, de la Torre and Burguete observed a
time-intermittency of the large Reynolds number turbulent state,
switching in between two metastable symmetry-breaking states
\cite{Torre2007,Burguete2009}. Such transitions with dynamical
symmetry breaking of mean patterns of the turbulent state are
observed in 2D turbulence simulations \cite{molenaar04,bouchet09},
where they have been claimed to have strong geophysical and
astrophysical relevance. They have also been observed and modeled
in dynamo regimes of a liquid-sodium turbulent flow
\cite{berhanu07,petrelis2008}.

Turbulent flows being intrinsically out-of-equilibrium systems,
there is a priori no reason to describe them using tools borrowed
from thermodynamics. However, in recent years, developments of
out-of-equilibrium thermodynamics led, in particular, to a number
of interesting applications to turbulence
\cite{ciliberto04,grenard08,monchaux08}. In this context, one may
wonder whether the observed turbulent transitions can be
interpreted in terms of phase transitions with a symmetry-breaking
or susceptibility divergence signature. In the above mentioned
works, while the Reynolds number appears as the natural control
parameter, the equivalent of an order parameter and its associated
susceptibility has not been clearly identified.

In this paper, we consider a von K\'{a}rm\'{a}n turbulent flow and
its averaged large-scale patterns. We introduce a susceptibility
to symmetry breaking for this flow and investigate its evolution
as $Re$ increases from $10^2$ to $10^6$ using Stereoscopic
Particle Image Velocimetry (SPIV). We show that this
susceptibility is as much as 50 times higher for the turbulent
flow at very high Reynolds number than for the laminar flow.
Furthermore, in a intermediate Reynolds number range ---$20\,000
\lesssim Re \lesssim 200\,000$---, we measure susceptibilities
values as high as 300 times the laminar value and conclude to the
existence, at a critical Reynolds number $Re_c = 40\,000 \pm
5\,000$, of a divergence of this susceptibility to symmetry
breaking. We also show that this phenomenon is associated with a
peak in the amplitude of the fluctuations of the flow
instantaneous symmetry. These observations complete the results of
reference \cite{cortet2010} with new complementary measurements at
$Re \lesssim Re_c$ and with a re-processing of all the
time-series. These new statistical analysis allow us to consider
not only the mean values of the distributions but also their
most-probable values. In particular, the divergence of the
susceptibility and the peak of fluctuations which were previously
found to occur at different values of $Re$ \cite{cortet2010} are
now observed at the same Reynolds number if we consider the
susceptibility based on most-probable values instead of mean
values. This result sets the threshold for a ``turbulent phase
transition'' at critical Reynolds number $Re_c = 40\,000 \pm
5\,000$. We relate the especially large fluctuations observed
close to the transition to time-intermittencies associated with an
ergodic exploration of a band of metastable states with
spontaneous symmetry breaking.

\section{System, flow, symmetries and order parameters}

\subsection{Experimental setup}

\begin{figure}
\centerline{\hspace{0.5cm}\includegraphics[width=12cm]{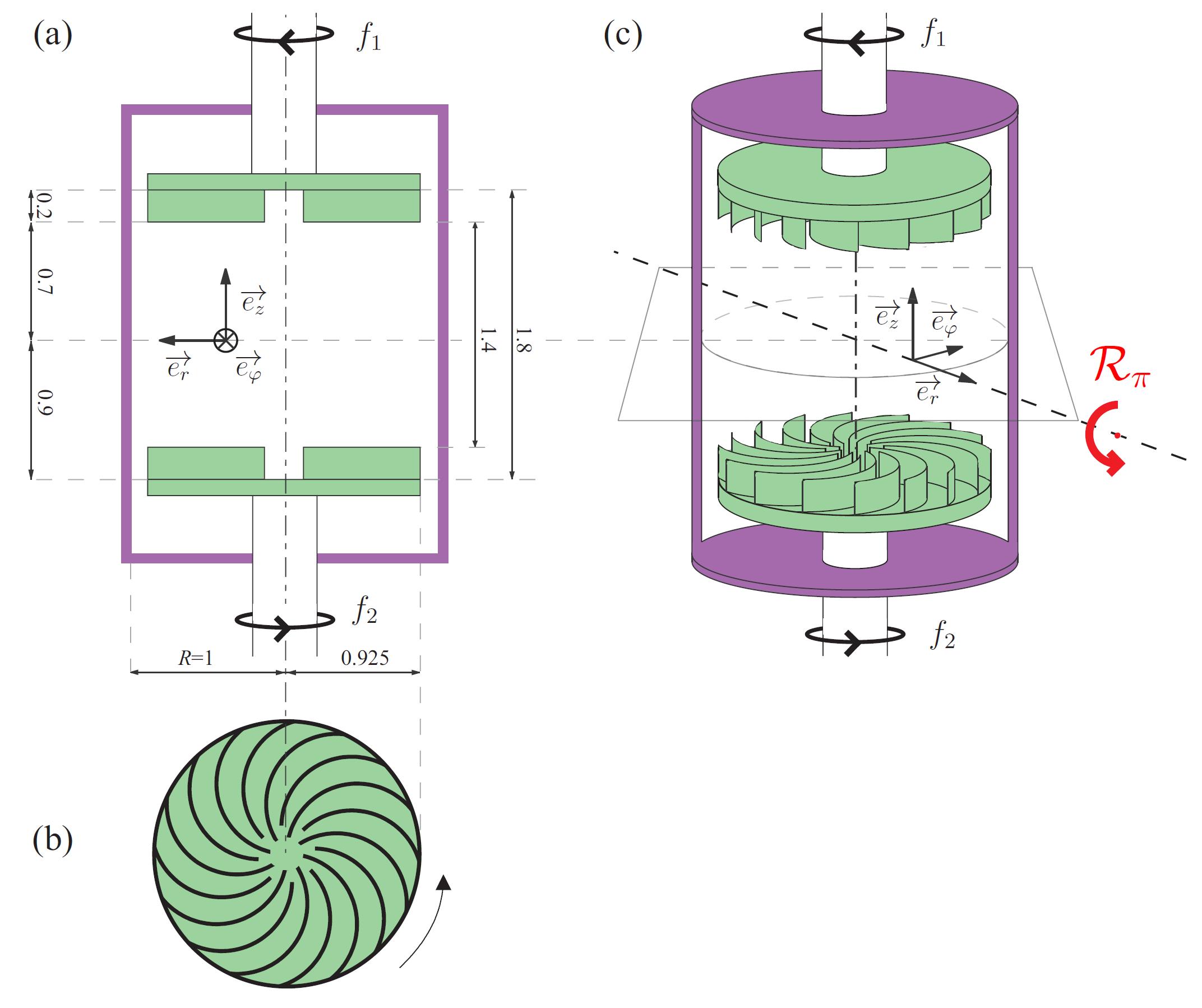}}
\caption{Schematic view of (a) the experimental setup and (b) the
impellers blade profile. The arrow indicates the rotation sense.
(c) 3D-view  and symmetry~: the system is symmetric with respect
to any $\cal{R}_\pi$-rotation of angle $\pi$ around any line in
the equatorial plane which crosses the rotation axis.}
\label{setup}
\end{figure}

Our experimental setup consists of a Plexiglas cylinder of radius
$R=100$ mm filled up with either water or water-glycerol mixtures
(\emph{cf.} figure \ref{setup}(a)). The fluid is mechanically
stirred by a pair of coaxial impellers rotating in opposite sense.
The impellers are flat disks of radius $0.925\,R$, fitted with 16
radial blades of height $0.2\,R$ and curvature radius $0.4625 \,R$
(\emph{cf.} figure \ref{setup}(b)). The disks inner surfaces are
$1.8\,R$ apart setting the axial distance between impellers from
blades to blades to $1.4\,R$. The impellers rotate, with the
convex face of the blades pushing the fluid forward, driven by two
independent brushless 1.8 kW motors. The rotation frequencies
$f_1$ and $f_2$ can be varied independently from $1$ to $12$ Hz,
speed servo loop control ensuring a precision of $2\permil$ for
the mean frequency $f=(f_1+f_2)/2$ and, for the relative frequency
difference $\theta=(f_1-f_2)/(f_1+f_2)$, an absolute precision of
$1 \times 10^{-3}$ and a stability along time better than $0.5
\times 10^{-3}$.

Velocity measurements are performed with a SPIV system provided by
DANTEC Dynamics. The cylinder is mounted inside a water or
glycerol filled square Plexiglas container in order to reduce
optical deformations. Two digital cameras are aiming at a meridian
plane of the flow through two perpendicular faces of the square
container providing 2D-maps of the three-components velocity
field. Correlation calculations are performed on $32\times 32$
pixels$^2$ windows with $50\%$ overlap. As a result, each velocity
is averaged on a $4.16 \times 4.16$ mm$^2$ window over the $1.5$
mm laser sheet thickness. The spatial resolution is $2.08$ mm. The
data provide the radial $u_r$, axial $u_z$ and azimuthal
$u_\varphi$ velocity components on a 95$\times$66 points grid
through time series of $400$ to $27\,000$ fields regularly
sampled, at frequencies from 1 to 15 Hz, depending on the
turbulence intensity and the related need for statistics.

The control parameters of the studied von K\'{a}rm\'{a}n flow are:
\begin{enumerate}[(i)] \item the Reynolds
number $Re= 2\pi f R^2/\nu$, where $\nu$ is the fluid viscosity,
which controls the intensity of turbulence and, \item the rotation
number $\theta=(f_1-f_2)/(f_1+f_2)$, which controls the asymmetry
of the forcing conditions.
\end{enumerate}

\subsection{System symmetry \textit{vs.} flow symmetry}

With exact counter-rotation of the impellers, \textit{i.e.} when
$\theta=0$, the experimental system (\emph{cf.} figure
\ref{setup}(a)) is symmetric with respect to any
$\cal{R}_\pi$-rotation exchanging the two impellers: the problem
conditions are invariant under $\pi$-rotation around any radial
axis passing through the center of the cylinder (\emph{cf.} figure
\ref{setup}(c)). The symmetry group for such experimental system
is $O(2)$ \cite{nore2003}. When the motors frequencies differ,
\textit{i.e.} when $\theta \ne 0$, the experimental system is no
longer $\cal{R}_\pi$-symmetric and the symmetry switches to the
$SO(2)$ group of rotations. However, the parameter $\theta$, when
small but non-zero, can be considered as a measure of the distance
to the exact $O(2)$ symmetry: the \emph{stricto sensu} $SO(2)$
system at small $\theta$ can be considered as a slightly broken
$O(2)$ system. Such perturbative approach of the symmetry breaking
has been successfully applied \cite{chossat93,porter2005} for the
1:2 spatial resonance (or $k-2k$ interaction mechanism) with
slightly broken reflection symmetry.

Driven at a given $\theta$ and a given $Re$, the system produces a
flow. This flow ---or at least its time-average at high $Re$---
may be ---or may not be--- $\cal{R}_\pi$-symmetric. The problem is
now to define a scalar quantity $S$, equivalent to an order
parameter, which can quantify the distance of the flow to the
$\cal{R}_\pi$-symmetry. The evolution of $S$ with $\theta$ will
measure the response of the flow to the symmetry of the forcing
and  $\chi=\partial S/\partial \theta$ will measure the
corresponding susceptibility. Identification of relevant $S$ can
be made by studying the flow topology in the laminar and steady
regime, at low Reynolds number, \emph{i.e.} for $Re \lesssim 200$.

\subsection{The von K\'{a}rm\'{a}n flow topology: the laminar case}

\begin{figure}
\centerline{\includegraphics[width=\textwidth]{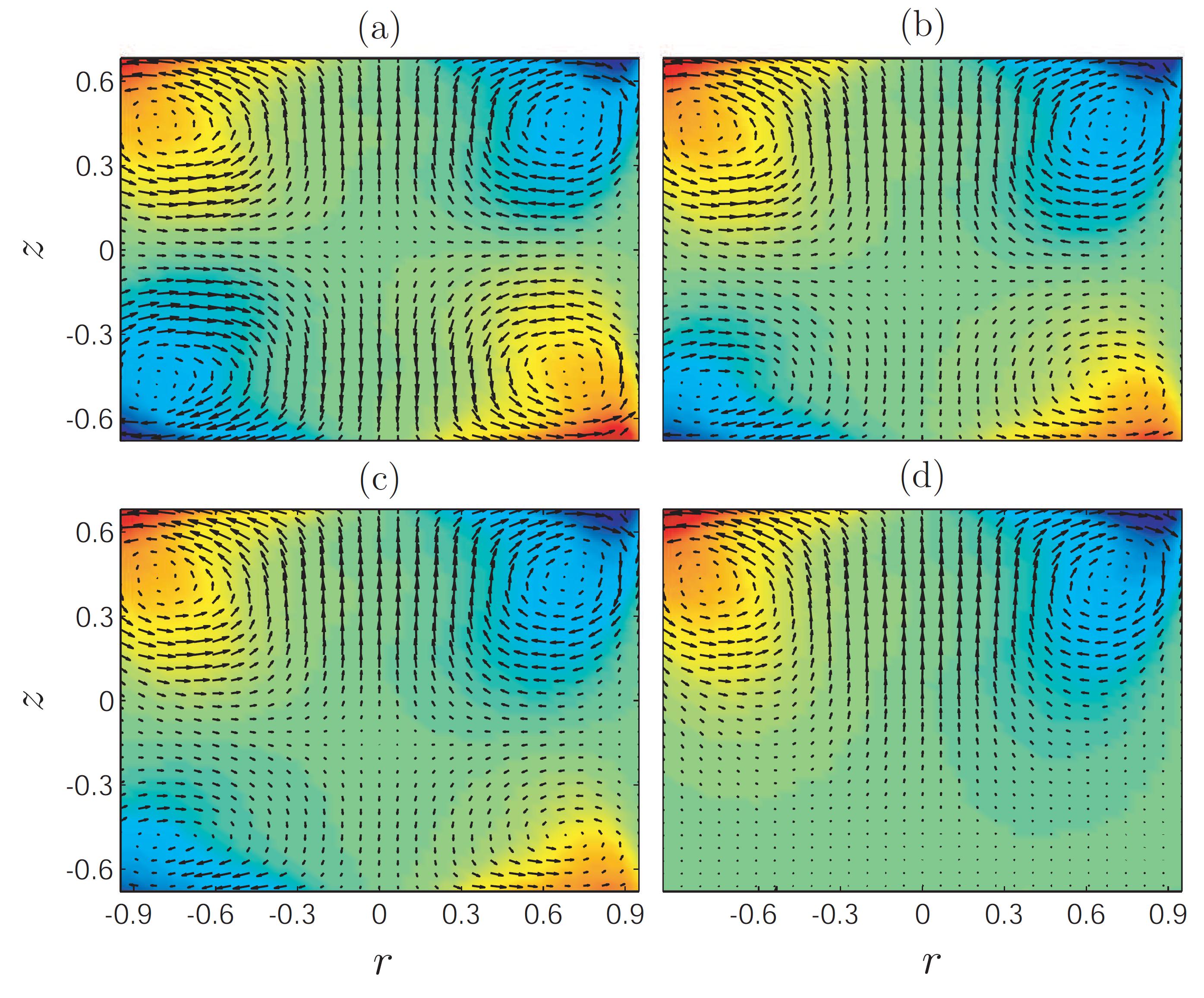}}
\caption{Maps of velocity fields in the meridian plane $(r,z)$
---the rotation axis is vertical at $r=0$ and the $r
\leftrightarrow -r$ symmetry of the maps reveals the flow
axisymmetry--- for the laminar ($Re=120 \pm 25$) von K\'arm\'an
flow for different values of the symmetry control parameter
$\theta$: (a) $\theta=0$, (b) $\theta=-0.11$, (c) $\theta=-0.19$,
(d) $\theta=-1$. The color maps, from blue to red (``jet''
colormap), the azimuthal velocity $u_\varphi$, whereas the arrows
map the $(u_r,u_z)$ field. The resolution of the fields has been
reduced by a factor 2 for better visibility. Positions $r$ and $z$
are given in units of vessel radius $R$.} \label{im_flow_lam}
\end{figure}

When $\theta=0$, the produced laminar flow is steady and composed
of two toric recirculation cells separated by an azimuthal shear
layer located at $z=0$ (\emph{cf.} figure \ref{im_flow_lam}(a))
reflecting the $\cal{R}_\pi$-symmetry of the system. Actually,
each impeller pumps the fluid located near the rotation axis into
its center before expelling it radially. This centrifugal pumping
creates a toric recirculation cell in front of each impeller. In
parallel, in each of the two toric cells, the fluid is rotating in
the azimuthal direction following the nearest impeller. Then, when
the motors frequencies differ, \textit{i.e.} when $\theta \ne 0$,
the shear layer moves towards the slowest impeller, breaking the
symmetry of the flow, but keeping the two-cells topology (\emph{cf.}
figures \ref{im_flow_lam}(b-d)). As $\theta$ tends to $\pm 1$, the
flow continuously tends to a one-cell topology \textit{i.e.} a
topology with a single toric recirculation cell.

\subsection{Identification of two order parameters and associated susceptibilities}

We will use two different order parameters to quantify the
distance of the flow to the $\cal{R}_\pi$-symmetry. First, we
consider a local quantity: the $z$-position of the shear layer of
the flow, \emph{i.e.} the position $z_s$ of the stagnation point
on the rotation axis\footnote{Practically, for an axisymmetric
flow, $z_s$ is measured as the axial position, $\psi(r=0,z_s)=0$,
of the zero iso-surface of the stream function $\psi(r,z)$ which
is defined through $(u_r,0,u_z)= \nabla\times (r^{-1}\psi\,{\bf
e}_\varphi)$ in cylindrical coordinates.}. The stagnation point
position $z_s$ is only defined when the flow has the particular
two-recirculation-cells axisymmetric structure and will not be
suitable ---because non-defined--- to monitor the distance of the
instantaneous flow to the $\cal{R}_\pi$-symmetry when it is
turbulent. For this purpose, we prefer a second order parameter, a
global quantity, the normalized and space-averaged angular
momentum $I(Re,\theta,t)$ :
\begin{equation}
I(t)=\frac{1}{\cal{V}}\int_{\cal{V}} rdrd\varphi dz\, \frac{r u_{\varphi}(t)}{\pi \, R^2 \,(f_1+f_2)}
\label{eq:I}
\end{equation}
where $\cal{V}$ is the volume of the flow. Practically, $I(t)$ is
computed from SPIV-data restricted to a meridian plane only but,
since azimuthal flow fluctuations are strong, time-average over
several impeller rotation periods ---statistically equivalent to
spatial azimuthal averaging--- estimate correctly the 3D-value of
$I(t)$ (see details in reference~\cite{cortet09}).

\begin{figure}
\centerline{\includegraphics[width=\textwidth]{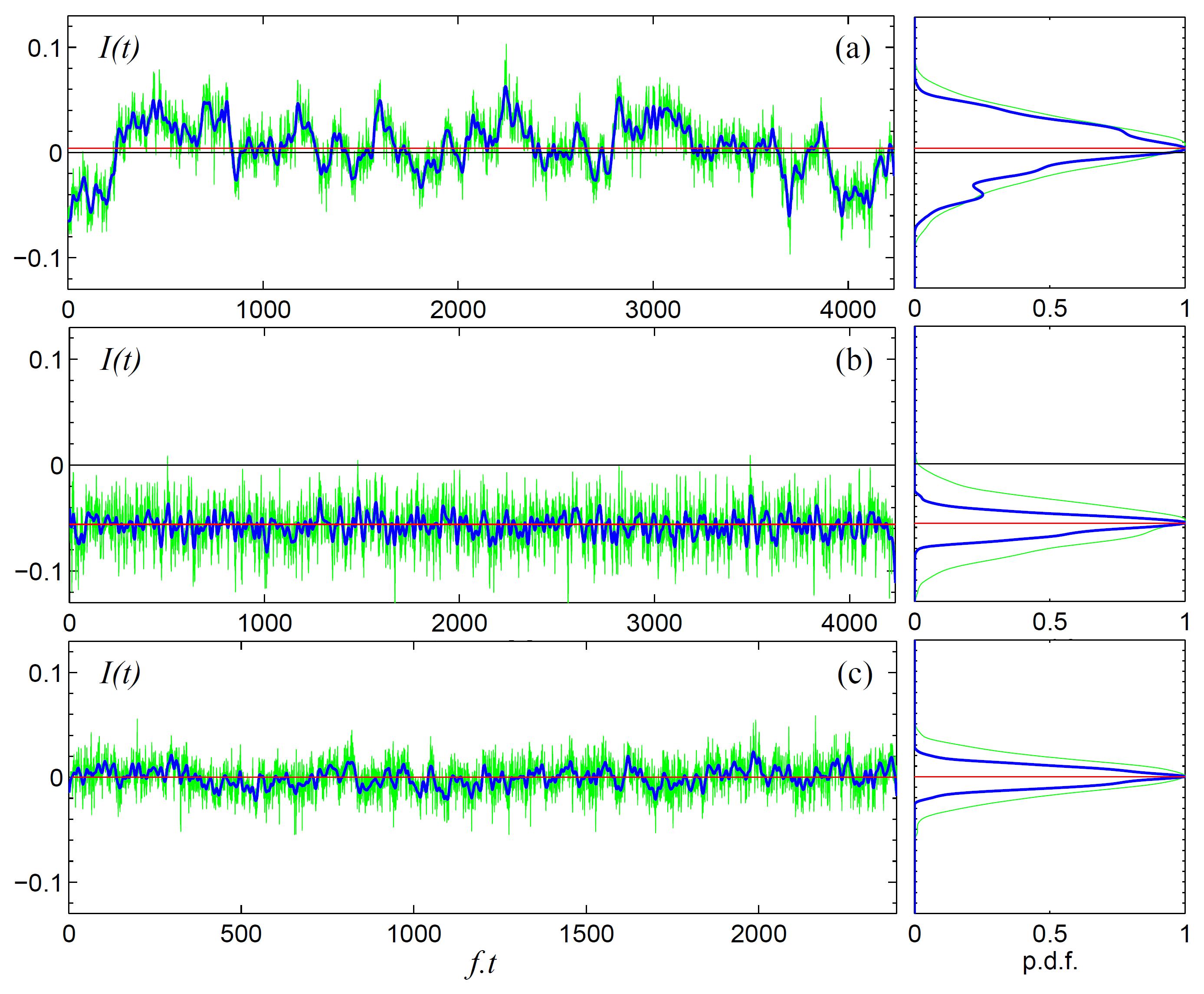}}
\caption{Global angular momentum $I(t)$ as a function of
non-dimensional time $f.t$ for experiments performed at
$Re=126\,700 \pm 200$ with (a) $\theta = 0 \pm 10^{-4}$ and (b)
$\theta=-0.0084$ and (c) at $Re=890\,000$ and $\theta=0$. Green
thin signals are SPIV data $I(t)$ sampled at 15 Hz and blue thick
signals correspond to low-pass filtered data $I_f(t)$.
Corresponding right figures shows the probability density
functions (PDF) of the signals with same color code. The red lines
indicate the most-probable values of $I_f(t)$.}
\label{series_temporelles_I}
\end{figure}

Examples of  time-series of $I(t)$  in turbulent regimes are
provided below in figure~\ref{series_temporelles_I} and reveal
that strong  spontaneous symmetry fluctuations can occur
(\emph{cf.} figure~\ref{series_temporelles_I}(a) for
$Re=126\,000$, $\theta = 0$). We assume that ergodicity holds,
meaning that the instantaneous turbulent flow is exploring along
time its energy landscape according to its statistical
probability. In this framework, the time average value
$\overline{I}$ of $I(t)$ is equivalent to a statistical mechanics
ensemble average providing the average is performed over a long
enough duration ---comparable to the averaging time needed to
measure $z_s$--- in order to correctly sample the slowest
time-scales. The quantity $I$ shows the major advantage of being,
contrary to $z_s$, a robust observable since it is defined for
each instantaneous velocity field whatever the structure of the
flow and therefore even for turbulent flows. When observing the
statistics of the fluctuations of $I$, we will sometimes encounter
non-Gaussian and even multi-peaked distributions. Therefore, we
will consider the time average $\overline{I}$, the most probable
value(s) $I_*$ and the fluctuations $I(t)$ of the global angular
momentum $I$ to characterize the mean, most-probable and
instantaneous symmetry of the flow.

Then, using order parameter $I$, we define two susceptibilities of
the flow to symmetry breaking $\chi_I$ and $\tilde{\chi}_I$, based
respectively on the mean and most-probable values of $I$, as:
\begin{equation}
\chi_I=\left.\frac{\partial \overline{I}}{\partial \theta}\right\vert_{\theta=0} \; \; \;  \rm{and} \; \; \;
\tilde{\chi_I}=\left.\frac{\partial {I_*}}{\partial \theta}\right\vert_{\theta=0}.
\label{eq:II}
\end{equation}
Another susceptibility may also be defined using $z_s$ as:
\begin{equation}
\chi_z=\left. \frac{dz_s}{d\theta}\right\vert_{\theta=0}.
\label{eq:III}
\end{equation}

\begin{figure}
\centerline{\includegraphics[width=\textwidth]{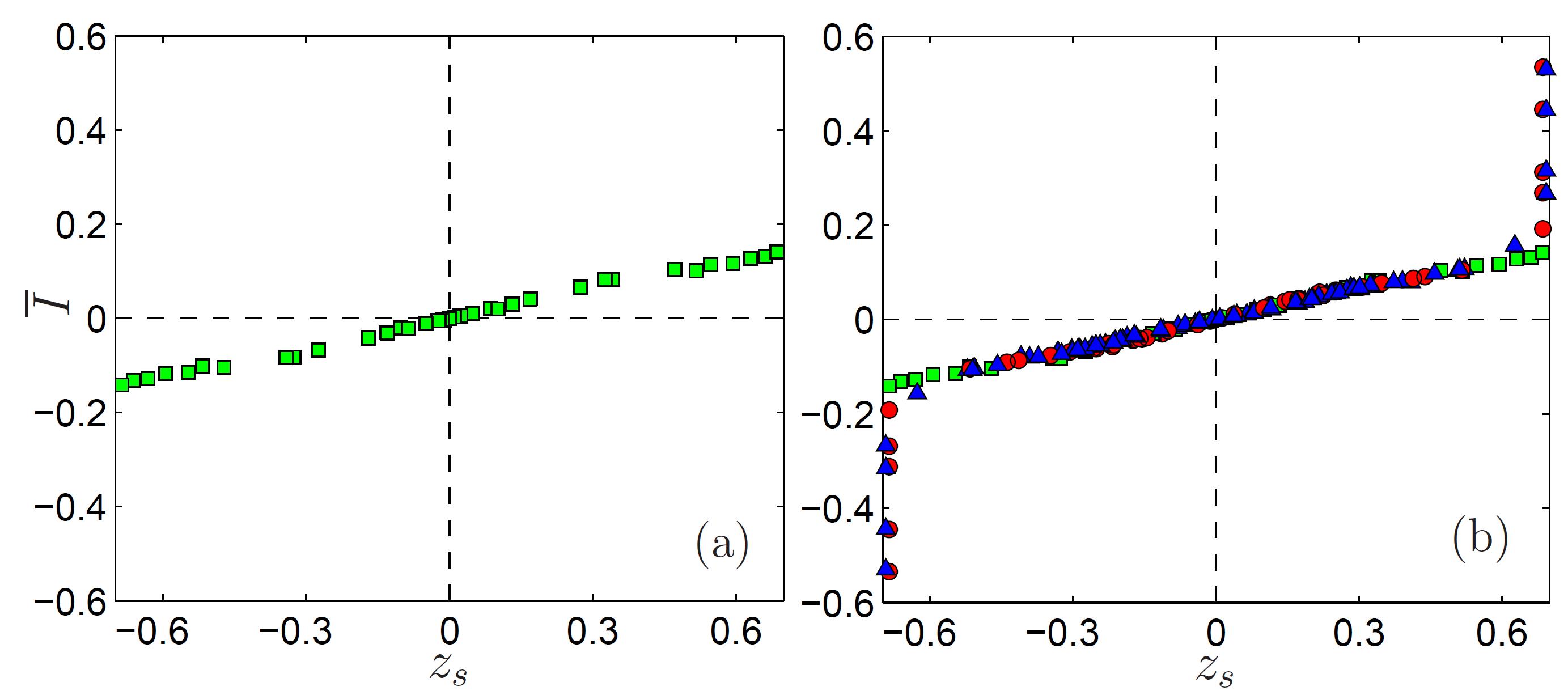}}
\caption{Experimental observation of proportionality between
time-averaged global angular momentum $\overline{I}$ and mean
shear layer altitude $z_s$ for (a) laminar flow $Re=120 \pm 25$
(green squares, $\overline{I}/z_s = 0.235 \pm 0.005$) and (b)
turbulent flows at $Re=67\,000$ (red circles, $\overline{I}/z_s =
0.227 \pm 0.005$) and $Re=890\,000$ (blue triangles,
$\overline{I}/z_s = 0.23 \pm 0.01$). Each data series is obtained
through plotting $\overline{I}(\theta,Re)$ as a function of
$z_s(\theta,Re)$ for varying $\theta$ and constant $Re$. At high
$Re$, this proportionality holds for $-0.1 \lesssim \theta
\lesssim 0.1$, \emph{i.e.} while the shear layer position $z_s$ is
not saturated to the impellers positions $\pm 0.7\, R$.}
\label{IvsZ}
\end{figure}

Finally, since our observations (\emph{cf.} figure \ref{IvsZ})
have shown that $\overline{I}$ is strictly proportional to the
mean altitude $z_s$ of the shear layer, with $\overline{I}/z_s =
0.23 \pm 0.01$ whatever $Re$, we may simply use the
susceptibilities $\chi= \chi_I = 0.23 \chi_z$ and $\tilde{\chi} =
\tilde{\chi}_I$. In the sequel, we investigate the influence of
turbulence on these order parameters ---$z_s$ and $I(t)$--- and
the associated susceptibilities as $Re$, and therefore turbulence
intensity, increases from $10^2$ to $10^6$.

\subsection{About the statistics of the time-series and high-frequency noise}

The statistics of the long time series of $I(t)$ obtained from
SPIV data represent a crucial step in our study but they are
limited by storage space and computing time. Our results are
partly based on the mean values $\overline{I}$ and its equivalent
$z_s$ computed on mean velocity fields and these lowest moments
are well converged for all the data presented. Since statistical
physics generally considers the most probable value of a
distribution rather than its mean value and since some signals
appear multivalued (\emph{cf.} figure
\ref{series_temporelles_I}(a) and figure \ref{serie_temp_I_39000}
below), we are also interested in measuring absolute and relative
most-probable values. However, the PDFs of the all raw signals
$I(t)$ are almost Gaussian and symmetric. Low-pass-filtering
applied on $I(t)$ with a typical $0.15$Hz-cutoff removes some
high-frequency noise, due in particular to the small scales of
turbulence, and leads to PDFs with relevant non-Gaussian shapes.
The best example is the high-resolution time-series corresponding
to the highest fluctuation-level measured at $Re=39\,000$,
presented in figure \ref{serie_temp_I_39000}. The most-probable
values are computed accordingly, but because of statistical
limitations, these value are less converged than mean values.

\subsection{Formal analogy with ferromagnetic systems}
\label{analogy}

In the non-fluctuating laminar case, when $\theta=0$,
$\overline{I}$ is strictly equal to zero due to the symmetry of
the flow. In contrast, as $\theta$ drifts away from $0$, the value
of the angular momentum $\overline{I}$ becomes more and more
remote from zero as the asymmetry of the flow grows. In the
turbulent regimes, the same behavior is observed but with higher
susceptibilities to symmetry breaking, \emph{i.e.} higher
sensitivity to $\theta$.

In such a framework, the Reynolds number ---or a function of it
\cite{castaing1996}--- is the equivalent of a statistical
temperature and we are entitled to propose the following formal
analogy between ferromagnetic systems and von K\'arm\'an systems:

\begin{itemize}
\item Order parameter: magnetization $M$ $\Leftrightarrow$ angular
momentum $I$; \item Symmetry breaking parameter: external applied
field $h$ $\Leftrightarrow$ relative driving asymmetry $\theta$;
\item Control parameter: temperature $T$ $\Leftrightarrow$
Reynolds number $Re$, or a function of it.
\end{itemize}

\section{Laminar and turbulent flow: influence of $\theta$ and $Re$}

\begin{figure}
\centerline{\includegraphics[width=\textwidth]{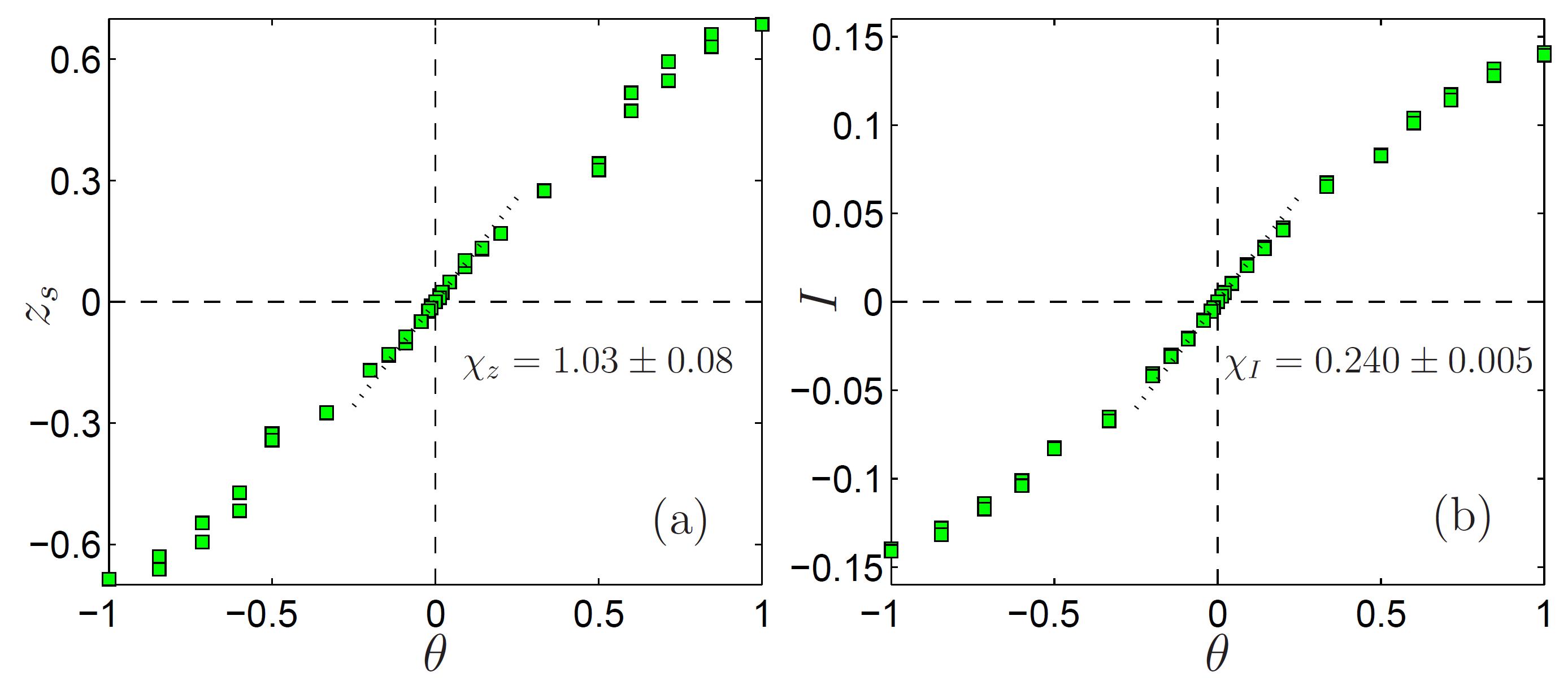}}
\caption{(a) axial position $z_s$ of the stagnation point and (b)
global angular momentum $I=\overline{I}$ as a function of $\theta$
for the laminar flows at $Re = 120 \pm 25$.}
\label{zstag_theta_lam}
\end{figure}

\subsection{The laminar susceptibility}

The evolutions of the symmetry parameters $z_s$ and $\overline{I}$
as a function of $\theta$ in the laminar flow at $Re=120 \pm 25$
are provided in figure \ref{zstag_theta_lam}. In this case
$\overline{I(t)}=I(t)$ and $z_s$ are defined for each
instantaneous field. We see that $z_s$ evolves almost linearly
with $\theta$ from $z_s=-0.7\,R$ (position of impeller 1) for
$\theta=-1$ to $z_s=0.7\,R$ (position of impeller 2) for
$\theta=1$, through $z_s=0$ for $\theta=0$ (\emph{cf.} figure
\ref{zstag_theta_lam}(a)). Variation of $I$ with $\theta$ is
similar. In the central region, the linearity is excellent and we
measure $\chi_z=1.03\pm 0.08$ and $\chi_I=0.240 \pm 0.005$.

\subsection{Flow topology : the turbulent case}

\begin{figure}
\centerline{\includegraphics[width=\textwidth]{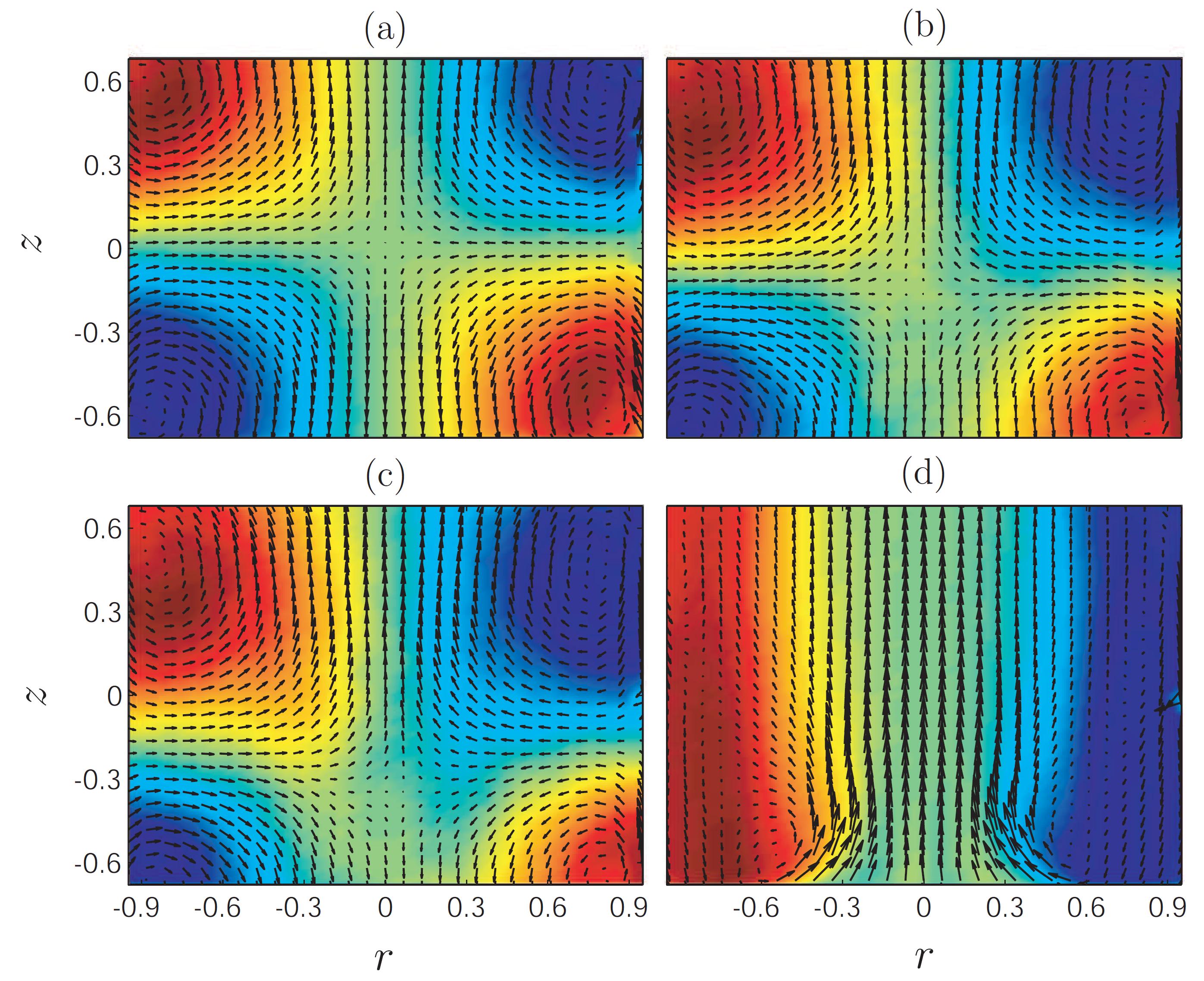}}
\caption{Maps of mean velocity fields of the turbulent von
K\'arm\'an flow at $Re=890\,000$ for different values of $\theta$:
(a) $\theta=0$, (b) $\theta=-0.0036$, (c) $\theta=-0.0147$, (d)
$\theta=-1$, with same layout as figure~\ref{im_flow_lam}. The $r
\leftrightarrow -r$ symmetry of the maps reveals that the
time-averaged mean fields are axisymmetric.}
\label{im_flow_turb_full}
\end{figure}

Increasing the Reynolds number, one expects to reach fully
developed turbulence around $Re=10\,000$ \cite{ravelet2008}. In
this turbulent regime and at $\theta=0$, the
$\cal{R}_\pi$-symmetry is broken for the instantaneous flow.
However, as usually observed for classical turbulence, this
symmetry is restored for the time-averaged flow (\emph{cf.} figure
\ref{im_flow_turb_full}(a)). Then, as in the case of the laminar
flow, when $\theta$ is varied, we observe the breaking of the
$\cal{R}_\pi$-symmetry of the turbulent mean flow, the structure
of which matches quite well the laminar flow topologies, however
with a higher sensitivity to $\theta$ (\emph{cf.} figures
\ref{im_flow_turb_full}(b-d)).

\subsection{The turbulent susceptibility}
\label{sec:turbsusc}

\begin{figure}
\centerline{\includegraphics[width=\textwidth]{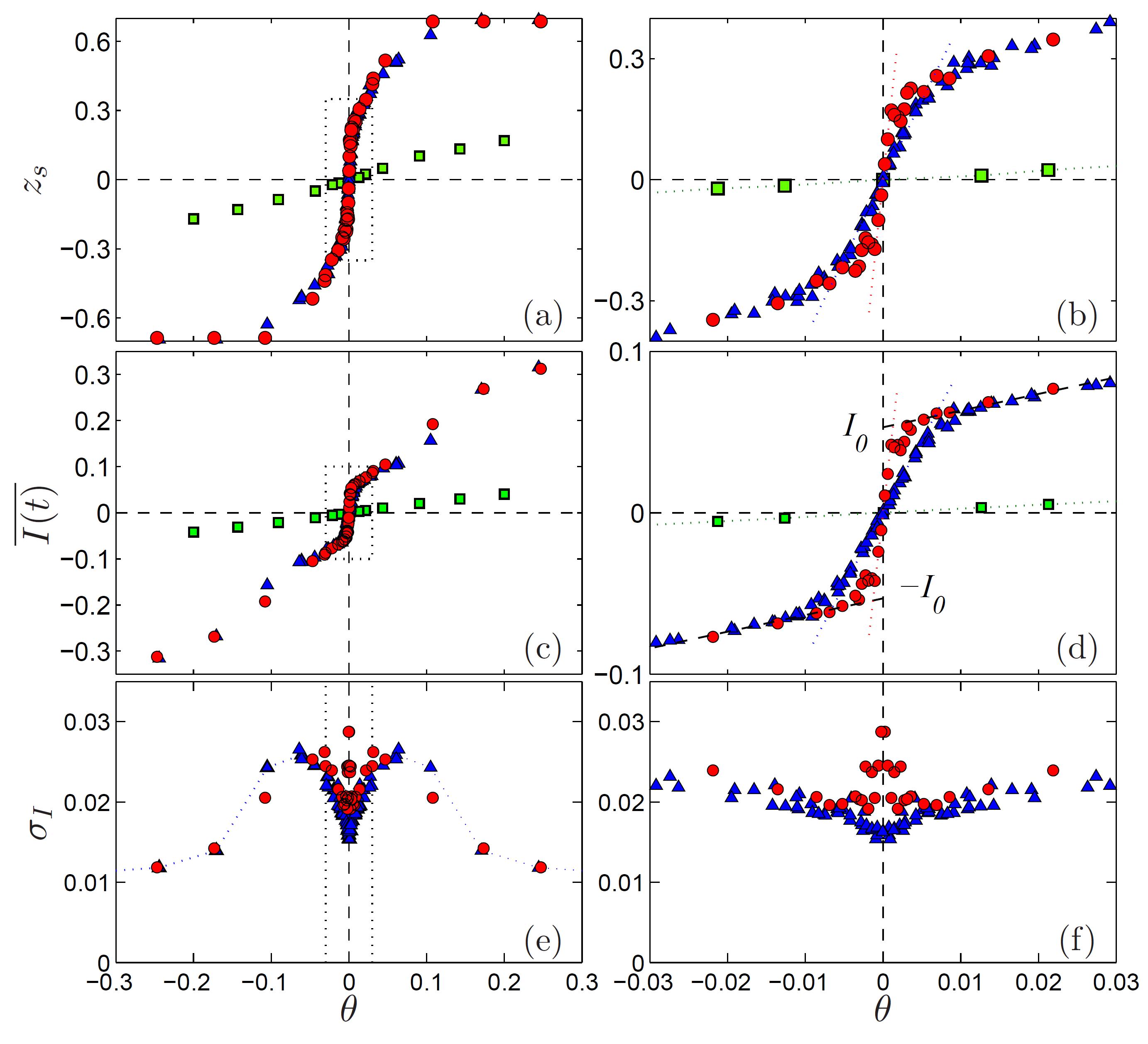}}
\caption{Turbulent regimes at $Re=67\,000 \pm 2\,000$ (red
circles) and $Re=890\,000 \pm 140\,000$ (blue triangles): axial
position of the stagnation point $z_s$ (a-b) and global angular
momentum $I$ ---mean $\overline{I(t)}$ (c-d) and standard error
$\sigma_I$ (e-f)--- as a function of $\theta$. Wide span
$-0.3<\theta<0.3$ on the left side (a,c,e) and $\times 10$ zoom
$-0.03<\theta<0.03$ on the right side (b,d,f). Data for the
laminar regime at $Re=120$ (green squares) from
figure~\ref{zstag_theta_lam} are also plotted for comparison. The
data have been symmetrized with respect to $\theta=0$. Dashed
lines in (d) indicates the tilted plateaus independent of $Re$ and
their extrapolations to $\pm I_0$.} \label{zstag_theta_turb}
\end{figure}

The main properties of $z_s$ and $I$ with respect to $\theta$ can
be observed in figure \ref{zstag_theta_turb} which combines the
measurements obtained at an intermediate Reynolds number
$Re=67\,000 \pm 2\,000$ (red circles) as well as at the highest
reachable Reynolds number in our experiment $Re=890\,000 \pm
140\,000$ (blue triangles). The left-hand-side figures give a
global view over a large span in $\theta$: at this scale the two
high-Reynolds-number series appear very close to each other and
show much higher susceptibilities than the laminar flow series
(green squares): turbulence enhances dramatically the sensitivity
of the flow to symmetry breaking. The turbulent susceptibilities
are at least 10 times larger than the laminar susceptibility since
$z_s$ saturates ---the shear layer disappears into one impeller---
for $|\theta|$ larger than a finite and small value $\theta_c =
0.10 \pm 0.02$, close to the value predicted in \cite{dijkstra}
for flat disks and 10 times smaller than in the laminar case where
the shear layer disappears asymptotically as $\theta\rightarrow
\pm 1$ (\emph{cf.} figure \ref{im_flow_lam} and
\ref{zstag_theta_lam}(a)). These disappearances of the shear layer
into impeller 1 or impeller 2 correspond to the transitions of the
flow topology from two recirculation cells ---at small
$|\theta|$--- to a single recirculation cell ---at large
$|\theta|$. In figure \ref{zstag_theta_turb}(e), we observe two
maxima for $\sigma_I$, the level of fluctuations of $I$, around
$\pm \theta_c$, which are characteristic of these transitions.
Such transitions have already been studied for other
counterrotating von K\'arm\'an flows in subcritical cases
\cite{ravelet2004,cortet09}.

The effect of the Reynolds number on turbulent flow responses is
revealed by the $\times 10$-zooms in $\theta$ on the
right-hand-side of figure \ref{zstag_theta_turb}. It is clear
that, in a narrow region $|\theta| \lesssim 0.01$, we encounter
two major differences between the $Re=67\,000$ and $Re=890\,000$
experiments:
\begin{itemize}
\item The susceptibilities are very different: $\chi_I=42 \pm 1$
at $Re=67\,000$ and $\chi_I=9 \pm 1$ at $Re=890\,000$ (\emph{cf.}
figure \ref{zstag_theta_turb}(b,d)), respectively $175$ and $37$
times the laminar value. \item The fluctuation level
$\sigma_I(\theta)$ presents a sharp and narrow peak at $\theta=0$
for $Re=67\,000$, but not for $Re=890\,000$ (\emph{cf.} figure
\ref{zstag_theta_turb}(f)).
\end{itemize}
This peak is typical of all data in the intermediate Reynolds
number range $20\,000 \lesssim Re \lesssim 200\,000$ where the
susceptibility is clearly higher than around the highest Reynolds
number reached $Re \sim 10^6$. Note that outside the active region
of high susceptibility and high fluctuations $|\theta| \lesssim
0.01$, there is no more $Re$-dependency for
$\overline{I}(\theta)$: in the medium range $0.01 \lesssim
|\theta| \lesssim 0.05$, not too close from the two-cells/one-cell
transition at $\pm \theta_c$, all curves collapse on tilted
plateaus which ---if extrapolated--- cross the $\theta=0$ axis at
$I_0 \simeq \pm 0.05$ whatever $Re$.

\subsection{Fluctuation level near $\theta=0$}

For all Reynolds numbers in the intermediate range, we observe a
sharp and narrow peak for the fluctuation level $\sigma_I(\theta)$
around $\theta=0$. This is illustrated by two time-series at
$Re=127\,000$, in figure \ref{series_temporelles_I}(a) and (b),
for $\theta \simeq 0$ and $\theta=0.0084$ which correspond
respectively to the top and the bottom of the fluctuation peak in
figure \ref{zstag_theta_turb}(f). The difference between the two
signals is striking and reveals that the fluctuation peak is due
to the presence non-Gaussian intermittencies. These
intermittencies have been observed in the range $|\theta| \lesssim
5\times10^{-3}$ exclusively. The $I(t)$ time-series at
$Re=890\,000$ and $\theta=0$, in figure
\ref{series_temporelles_I}(c), corresponds however to the minimum
fluctuation level of figure \ref{zstag_theta_turb}(f): the
corresponding PDF is narrow, symmetric and has a single central
maximum.

\subsection{Mean \emph{vs.} most-probable values for highly intermittent regimes}

\begin{figure}
\centerline{\includegraphics[width=\textwidth]{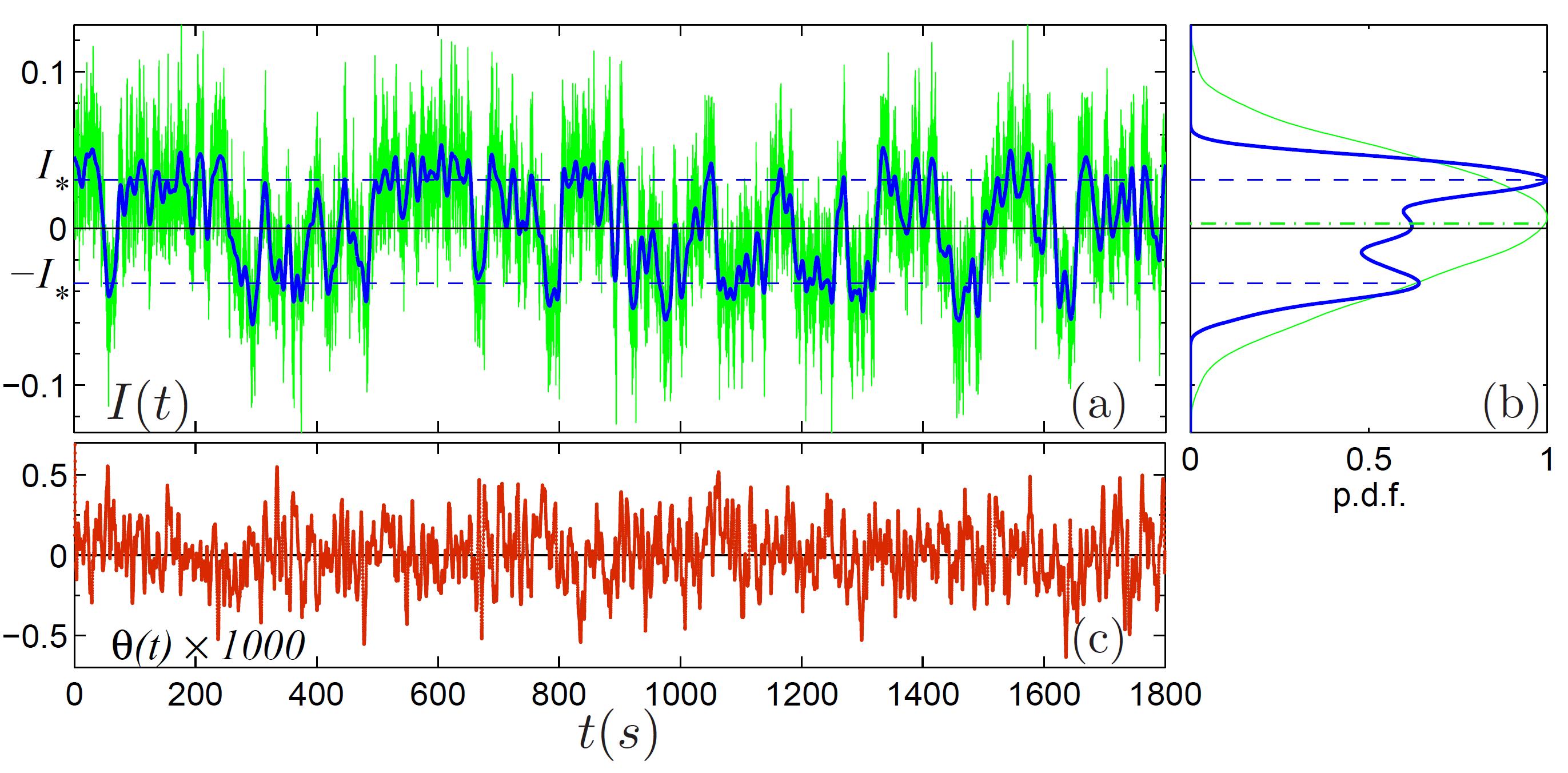}}
\caption{(a) Global angular momentum $I(t)$ for an experiment
performed at $Re=39\,000$ for $\theta=\overline{\theta(t)}=0$. The
green thin line is SPIV data $I(t)$ sampled at 15 Hz and the blue
thick line corresponds to 0.15 Hz low-pass filtered data $I_f(t)$.
(b) Probability density functions (PDF) with the same color code.
Blue dashed lines indicate the two most-probable values $\pm I_*$
of $I_f(t)$ and green dash-dot lines indicate the mean value
$\overline{I}$. (c) Time-series of $\theta(t)$ as produced by the
motor regulation in unit of $10^{-3}$: the correlation with $I(t)$
is negligible (see text).} \label{serie_temp_I_39000}
\end{figure}

In the above paragraphs and figures we have presented the
evolution of mean values of $I(t)$ with $\theta$ for different
$Re$. Globally, most-probable values differ from mean values only
when the susceptibility is high and are, in absolute value,
greater than mean values. This difference is especially noticeable
for $Re$ around $40\,000$. The data set which corresponds to the
maximum fluctuation level is presented in figure
\ref{serie_temp_I_39000} where $\overline{I} \simeq 0$. It shows
strong temporal intermittency between states of different angular
momentum values revealed by the multi-peak PDF.

\begin{figure}
\centerline{\includegraphics[width=0.8\textwidth]{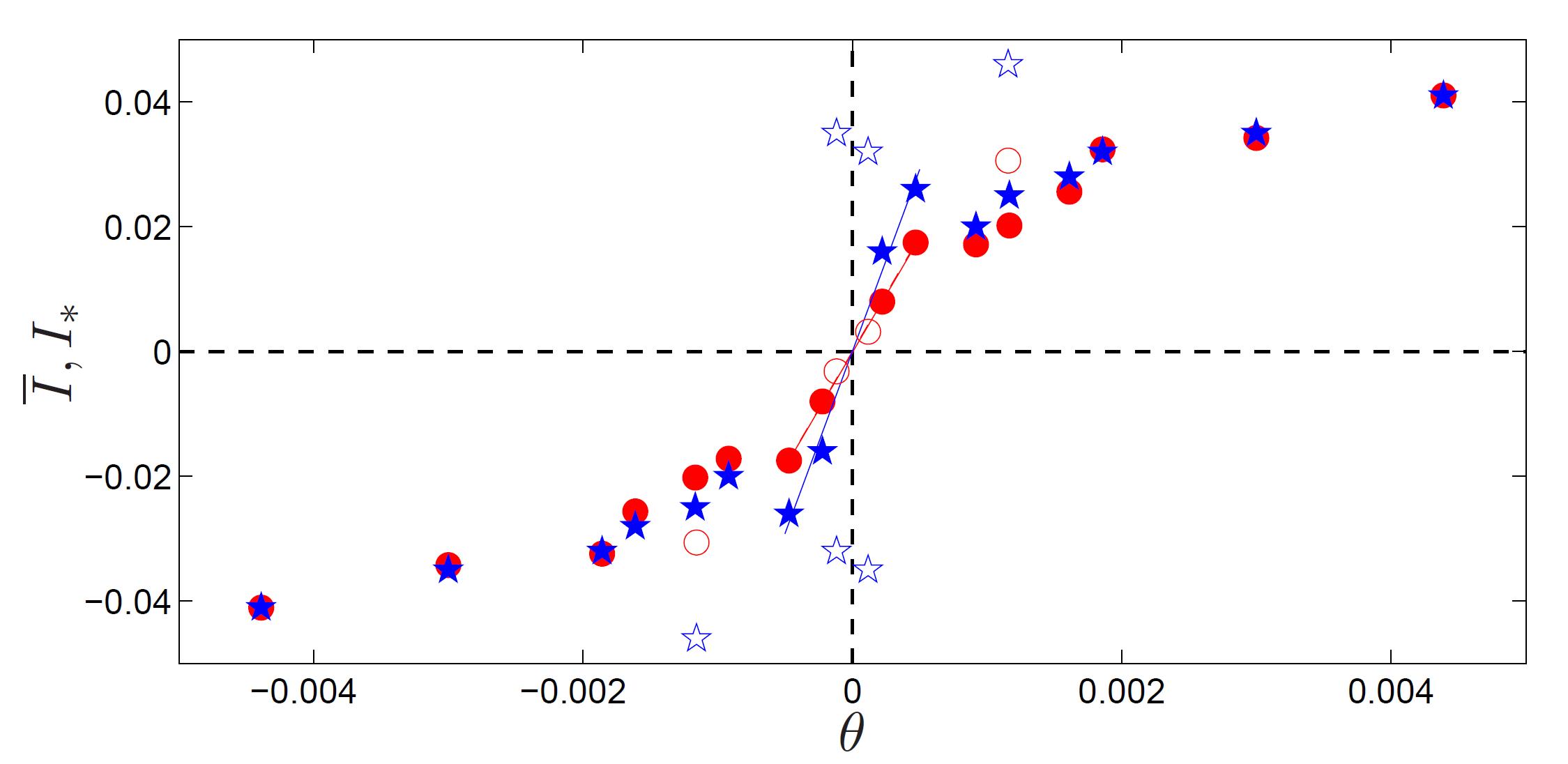}}
\caption{Mean $\overline{I}$ (red circles) and most-probable $I_*$
(blue stars) values of the global angular momentum $I(t)$ for $Re
\sim 40\,000$. Close symbols come from a set of
moderate-resolution data at $Re = 37\,000 \pm 2\,000$ ($f=12$Hz,
$1\,200$ SPIV fields acquired at $1.7$Hz), whereas open symbols
come from high-resolution time-series at $Re = 41\,000 \pm 2\,000$
($f=8$Hz, $9\,000$ or $27\,000$ SPIV fields acquired at $15$Hz as,
e.g., in figure \ref{serie_temp_I_39000}) acquired in a different
glycerol-water mixture. The data have been symmetrized with
respect to $\theta=0$ and, for the data point at $\theta \simeq 0$
of the second series, we plot the two most-probable values $\sim
\pm I_*$ corresponding to the two peaks of the PDFs as in figure
\ref{serie_temp_I_39000}(b).} \label{most-prob}
\end{figure}

The most-probable values of $I(t)$ are plotted in figure
\ref{most-prob}, with respect to $\theta$ very close to
$\theta=0$. We observe that for $Re = 37\,000 \pm 2\,000$, the
slope $\tilde{\chi}$ for most-probable values is approximately
twice the slope $\chi$ for mean values at $\theta=0$. We also
reported points obtained from three high-resolution time-series at
$Re=39\,000$ (figure \ref{serie_temp_I_39000}) and $43\,000$. For
these data sets, we plot the first two most-probable values $\pm
I_*$ of $I(t)$ corresponding to the two almost symmetric peaks of
the PDFs with are exchanging their relative importance around
$\theta=0$. These values are significantly higher than those of
the $Re = 37\,000$-data series. This can be due to either the
proximity of the critical value $Re_c$ or to the limited
statistics of the $Re = 37\,000$-data series. Anyhow, these
results suggest that the corresponding susceptibility
$\tilde{\chi}$ is diverging at $Re=41\,000 \pm 2\,000$ as it will
be discussed in the next section.

\subsection{Divergence of the susceptibility at $Re \simeq 40\,000$}

From the above results it is clear that the effect of the Reynolds
number is to be studied for $\theta=0$. The present section deals
with the susceptibilities and the fluctuation level at $\theta=0$
and for $Re$ covering the whole study range $10^2 - 10^6$.

\begin{figure}
\centerline{\includegraphics[height=0.4\textwidth]{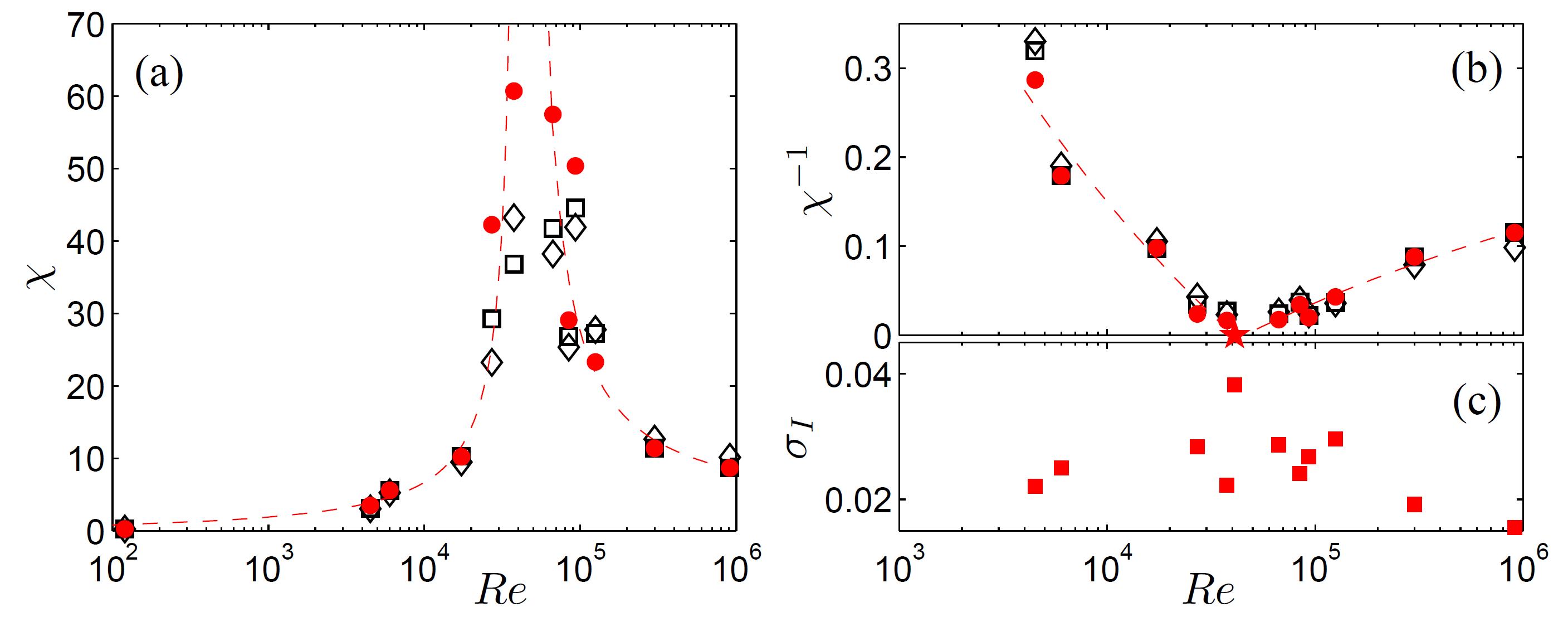}}
\caption{Reynolds number dependence of the symmetry-breaking
susceptibility of the von K\'arm\'an flow at $\theta=0$ (a) and
its inverse (b). Black open symbols correspond to susceptibility
$\chi = \chi_I = 0.23 \chi_z$ obtained from mean values
$\overline{I}$ (squares) and $z_s$ (diamond); red full circles
stands for $\tilde{\chi}_I$ obtained from most-probable values
$I_*$ of $I$. The red full star corresponds to the region
($Re=41000 \pm 2000$) where infinite susceptibility is inferred
from the strong multistability of the signal (\emph{cf.} figures
\ref{serie_temp_I_39000} and \ref{most-prob}). Dashed lines are
fits of the most-probables values above and below the star as
linear functions of $1/\log(Re)$ as proposed in the discussion of
section \ref{sec:critical}. (c) Standard deviation $\sigma_I$ of
the unfiltered global angular momentum $I(t)$ at $\theta=0$.}
\label{zstag_reynolds}
\end{figure}

In figure \ref{zstag_reynolds}(a), we plot susceptibility
measurements ---$\chi$ and $\tilde{\chi}$, \emph{cf.} equation
\ref{eq:II}--- as a function of $Re$. Figure
\ref{zstag_reynolds}(b) shows the inverses of these values and
figure \ref{zstag_reynolds}(c) the corresponding standard
deviation $\sigma_I$ of $I(t)$. The susceptibility $\tilde{\chi}$
based on the most probable values $I_*$ of $I(t)$ and the
fluctuations of $I(t)$ are both maximum for $Re \simeq 40\,000$.
However, as reported earlier \cite{cortet2010}, the susceptibility
$\chi$ based on the mean values of $I(t)$ reaches a maximum for
higher $Re \simeq 90\,000$. We also note that $\tilde{\chi}
\apprge \chi$, whatever $Re$, and that the highest measured
susceptibilities are of the order of the highest measurable values
considering the $\theta$-precision of our setup.

All the above results suggest a divergence of $\tilde{\chi}(Re)$
which clearly appears with the fits in figures
\ref{zstag_reynolds}(a) and (b): $\tilde{\chi}(Re)$ diverges at
$Re = Re_c = 40\,000 \pm 5\,000$ with critical exponent $-1$
[illustrated in figure \ref{zstag_reynolds} with respect to
$1/\log(Re)$ (see discussion in section \ref{sec:critical})] and
resembles a classical magnetic susceptibility divergence at the
ferromagnetic/paramagnetic phase transition. In the next section,
we investigate further this transition by looking at the
equivalent of the spontaneous magnetization of magnets,
\emph{i.e.} how, around $Re_c$, the angular momentum $I(t)$
spontaneously and dynamically transits between different finite
values.

\section{Dynamic multistability near $Re_c$ and spontaneous ``momentization''}

\subsection{Observation of the dynamics at $Re=39\,000$}

Close to the susceptibility divergence, we observe a complex
dynamics for the global angular momentum $I(t)$, as illustrated in
figure \ref{serie_temp_I_39000} at $Re=39\,000$ and $\theta=0$.
Indeed, one observes that $I(t)$ does not just fluctuate randomly
around zero ---its mean value--- but shows a tendency to lock for
some times, preferentially around $\pm I_*$ with $I_*$ of the
order of $0.03$-$0.04$, estimated from the main peaks of the PDF
of the filtered signal $I_f(t)$ (\emph{cf.} figure
\ref{serie_temp_I_39000}(b)).

\begin{figure}
\centerline{\includegraphics[width=\textwidth]{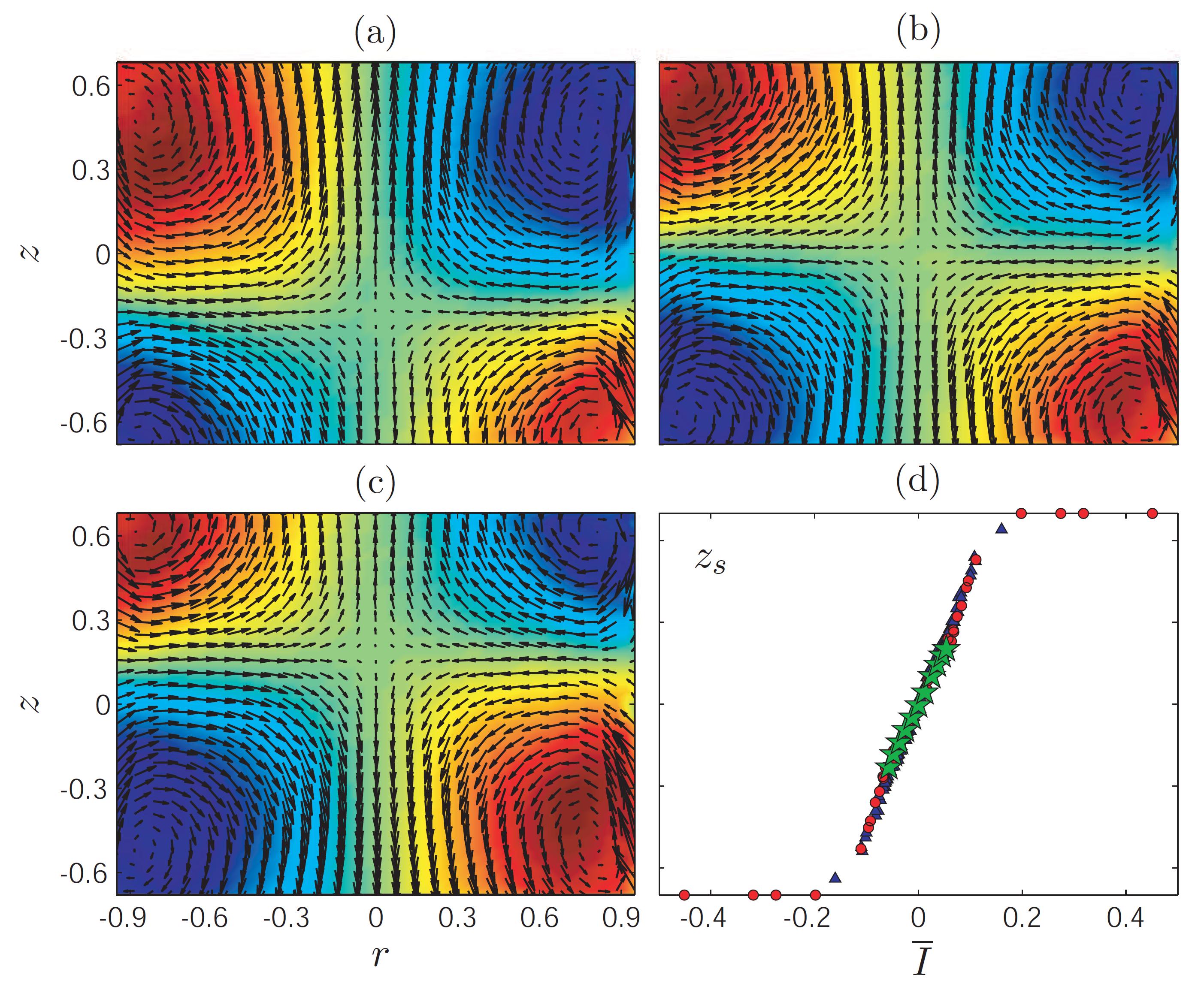}}
\caption{Bin-averaging of the velocity field with respect to
$I(t)$ for $Re=39\,000$ and $\theta=0$ (same data series as in
figure \ref{serie_temp_I_39000}). Three bin-averaged velocity
fields are presented for bins:  (a) $-0.075 < I_f(t) < -0.050$,
(b) $-0.0125 < I_f(t) < 0.0125$ and (c) $0.0375 < I_f(t) <
0.0625$. (d) Corresponding stagnation point axial position $z_s$
measured for 11 such bin-averaged fields for $I$ between $-0.075$
and $+0.075$ (green stars), superimposed to regular time-averaged
data for varying $\theta$ at $Re= 67\,000$ (red circles) and
$Re=890\,000$ (blue triangles) as in figure~\ref{IvsZ}(b).}
\label{cond_field}
\end{figure}

In order to understand the nature of the fluctuations of $I(t)$,
we have first checked that $I(t)$ is uncorrelated with the control
parameter $\theta(t)$ (correlation coefficient $C(I,\theta)=
\overline{I(t)\,\theta(t)}/\sigma_I\,\sigma_\theta < 0.02$,
\emph{cf.} figures \ref{serie_temp_I_39000}(a,c)) and thus reveals
only spontaneous fluctuations. Actually, $I(t)$ fluctuates with
two separate time scales: \emph{(i)} fast fluctuations related to
``traditional'' small-scale turbulence and \emph{(ii)}
intermittencies corresponding to residence time of typically few
tens of seconds and certainly due to the proximity of the
susceptibility divergence. If one performs a time-average of the
velocity field over one of these intermittent periods, one obtains
a time localized ``mean'' flow, analogous to what is obtained for
true mean flows when $\theta\neq0$ as presented in figures
\ref{im_flow_turb_full}(b-c), and with its own level of
spontaneous symmetry breaking. To go one step further with the
correspondence between $I(t)$ and the amount of symmetry breaking
or the position of the shear layer, we have computed bin-averages
of the velocity field for a time-series $I(t)$ in 11 bins ranging
from $-0.075$ to $+0.075$. In figures \ref{cond_field}(a-c), we
have plotted three examples of such bin-averaged fields. The
pattern is $\cal{R}_\pi$-symmetric for the bin around zero
(\emph{cf.} figure \ref{cond_field}(b)) and non-symmetric
otherwise (\emph{cf.} figures \ref{cond_field}(a) and (c)). From
the resulting series of conditionally-averaged velocity fields we
measure the stagnation point axial position $z_s$ and report it as
a function of the mean bin value (\emph{cf.} figure
\ref{cond_field}(d)). There is a very good matching between these
bin-averaged data and the time-averaged data already presented in
figure \ref{IvsZ} for different $\theta$: beyond the
proportionality between $z_s$ and $\overline{I}$, $I(t)$ can thus
definitively be used as a quantitative instantaneous symmetry
measurement. For example, the two symmetrical most-probable values
$\pm I_*$ of $I(t)$ correspond typically to $|z_s| \simeq 0.2\,R$.

From the above observations, we can conclude that, very close to
$Re_c$, the turbulent flow explores a band of metastable symmetry
breaking patterns evidenced by $-I_* \lesssim I_f(t) \lesssim
+I_*$, where $I_f(t)$ is the $0.15$Hz low-pass filtered value of
$I(t)$, the three most-visited states being $-I_*$, $0$, and
$+I_*$. Farther from $Re_c$, signals and PDFs also reveal
coexisting preferred values for $I(t)$ when $\theta$ is very
small.

\subsection{The ``momentization''}

The dynamical spontaneous symmetry breaking observed  in our
experiment can be seen as the coexistence of
multistable/metastable steady states of different symmetry which
are explored dynamically along time. By analogy with magnetization
we propose to describe this as successive ``momentization''
patterns observed in time, thanks to ergodicity which arises from
the very high level of fluctuations.

With respect to magnetism, the main difference is that the
spontaneous momentization occurs along time and mainly very close
to the critical point. Furthermore, owing to the statistical
limitations, we cannot conclude if momentization occurs only on
one side of the transition ---as it is for the ferro/para
transition--- or on both sides.

\subsection{Statistics of the dynamic multistability}
\label{stat}

The metastable/multistable states of finite lifetime ($\tau$)
observed in time-series such as in figure
\ref{serie_temp_I_39000}(a) can be characterized by their angular
momentum averaged over their lifetime and noted
$\overline{I}^\tau$, to avoid confusion with the full-signal mean
value $\overline{I}$. The $\overline{I}^\tau$ are distributed
along the angular momentum axis and we note ${\cal{I}}(Re,\theta)$
the $I$-interval containing the distribution for experiments at
$(Re,\theta)$.

Multistability is systematically observed in the high
susceptibility zone limited by $Re \in {\cal{I}}_{Re} \sim
[20\,000,200\,000]$ and $\theta \in {\cal{I}}_{\theta} \sim
[-0.01,+0.01]$. Whatever $Re$ and $\theta$ in ${\cal{I}}_{Re}$ and
${\cal{I}}_{\theta}$, we observe that ${\cal{I}}(Re,\theta)$ is
confined in $[-I_0,+I_0]$ ($I_0$ has been defined at the end of
section \ref{sec:turbsusc} and is shown in figure
\ref{zstag_theta_turb}). When $Re \simeq Re_c$, the interval
${\cal{I}}(Re,\theta) \simeq [-I_*,+I_*]$ is symmetric and close
to $[-I_0,+I_0]$ (\emph{cf.} figure \ref{serie_temp_I_39000}): the
whole band is almost populated. However, farther from $Re_c$, the
interval ${\cal{I}}(Re,\theta)$ is smaller and evolves inside
$[-I_0,+I_0]$ with $\theta$.

\section{Complementary investigation about the azimuthal structure of the flow}
\label{Eckhaus}

Before discussing our results in the next section, it is important
to note that all the above results concern 2D-velocity-fields in a
meridian plane: the azimuthal spatial structure is ignored by our
2D-SPIV process since the azimuthal patterns ---moving and
fluctuating--- are averaged along time and not resolved in space
at all. In this section, we investigate some preliminary
experimental results about the 3D-structure of the flow, gathered
by visual observation with bubble seeding and white light
\cite{ravelet2008,cortet09}. With a high density of bubbles we
mostly see the patterns on the outer cylinder which are the radial
vortices due to the shear-layer destabilization. In practice, we
experience that, even in the flow is turbulent, the azimuthal
number $m$ of vortices can be a well defined integer over some
ranges of Reynolds number and/or some periods of time. So, exactly
as for small dynamical systems, changes in $m$ can reveal an
Eckhaus instability for the 3D-mean pattern \footnote{We use here
the word ``mean'' not for the 2D- or axisymmetric-time-average, as
everywhere in the present article: here, ``3D-mean'' involves time
averaging in a rotating frame following the azimuthal pattern.
Such image processing will be a challenge, implying movies of the
whole system \cite{prigent00} or full 3D velocimetry.}.

This dynamics has been observed for $22\,000 \apprle Re \apprle
300\,000$. Despite the vortices are strongly fluctuating, we can
report the following observations of the average number $m$ of
vortices around the perimeter of the shear layer at $\theta=0$:
\begin{itemize}
\item $m=3$ for $22\,000 \apprle Re \apprle 47\,000$,
\item $m=4$ for $86\,000 \apprle Re \apprle 300\,000$.
\end{itemize}
\noindent Between these two regions, the fluctuations are
generally too strong to draw precise conclusions, but at least a
clear observation of $m=4$ has been made once at $Re \simeq
51\,000$. So, we conclude that some kind of $m=3 \leftrightarrow
m=4$ Eckhaus transition occurs between $Re=47\,000 \sim Re_c$ and
$Re=86\,000 \sim 2 Re_c$, in the region of very high
susceptibility.

\section{Discussion}

We have reported the experimental study, for a von K\'{a}rm\'{a}n
swirling flow, of \emph{(i)} the response of the mean flow to a
continuous breaking of the $\cal{R}_\pi$-symmetry of the system
and \emph{(ii)} the spontaneous symmetry fluctuations of the
instantaneous flow, from the laminar regime at $Re \sim 10^2$ to
the highly turbulent regime at $Re \sim 10^6$. The divergence of
the susceptibility $\tilde{\chi}(Re)$ for $Re = Re_c = 40\,000 \pm
5\,000$ reveals the existence of a phase transition or a
bifurcation. In the following, we discuss the implications of this
phase transition for turbulence, the critical behaviors and the
analogy with the para-ferromagnetic transition.

\subsection{A Reynolds-dependent turbulence}

This new transition makes some non-dimensional characteristic
quantities of the flow still $Re$-dependent at very high $Re$. The
turbulence is thus definitively not fully developed in this closed
turbulent flow, contrary to the most common observations in this
Reynolds number range. In other closed flows, the literature
reports some examples of transitions occurring at Reynolds number
of the order of $10^5$, a decade in $Re$ beyond where the
turbulence already looks like fully developed. First, in various
type of turbulent fluid experiments, several authors report
transitions and symmetry breaking at high Reynolds number
(\emph{cf.}, e.g., references
\cite{chilla2004,mujica2006,gibert2009,lohse2010}). Moreover, in a
similar von K\'arm\'an flow,  de la Torre and Burguete
\cite{Torre2007,Burguete2009} observed a bistability between two
broken $O(2)$-symmetry states with $z_s/R= \pm 0.19$. Although the
slow dynamics is very different from what is reported here, both
experiment are likely to belong to a common framework (see below).
Finally, in a liquid helium von K\'arm\'an flow, Tabeling
\textit{et al.} report a local peak in the flatness of velocity
derivative around $R_\lambda=700$ ---corresponding to $Re=2 \times
10^5$ \cite{tabeling96}. An interpretation in terms of a
second-order phase transition was proposed \cite{tabeling02}.
Velocity derivatives cannot be measured in our experiment because
of the low spatial resolution of our SPIV system. However, if one
could establish that the observations of Tabeling \textit{et al.}
are indeed connected with ours, then our work would prove that
their transition is related to the global structure of the mean
flow and thus does not result from the breakdown of small-scale
vortical structures as proposed in reference~\cite{tabeling02}.

\subsection{Critical behaviors and analogy with magnetism}
\label{sec:critical}

\subsubsection{The control parameter: a function of $\log Re$ ---\label{sec:Re}}

Since our study covers 4 orders of magnitude in Reynolds number,
it is natural to acquire and plot results along a logarithmic
scale. However, we could search for critical behaviors with either
$Re$ or $\log Re$ as control parameter. Indeed, critical behaviors
are asymptotic behaviors around a given threshold $Re_c$ and
therefore both $(Re-Re_c)$ and $(\log Re - \log Re_c)$ are
mathematically equivalent. However, it is very common in
non-linear physics, that asymptotic behaviors or normal forms
correctly describe the dynamics, not only asymptotically, but over
finite ---and even large--- ranges around thresholds using one
particular control parameter. This is exactly what happens here
and because of the quality of the $1/\tilde{\chi}(Re)$ fit in
figure~\ref{zstag_reynolds}(b), we conclude that a relevant
control parameter for this critical phenomenon can be $\log Re$ or
a gentle function of it.

Within the formal turbulence/magnetism analogy, we
actually use Castaing's proposition \cite{castaing1996} for a
temperature of turbulent flows which is $T \sim 1/\log Re$.

\subsubsection{Critical behavior of the susceptibility ---}

The divergence appears clearly in figure \ref{zstag_reynolds}(b)
where $1/\tilde{\chi}(Re)$ goes to zero at $Re = Re_c = 40\,000
\pm 5\,000$ in a way that can be fitted with a $-1$ critical
exponent by:
\begin{equation}
\tilde{\chi} \propto \left \vert \frac{1}{\log Re}- \frac{1}{\log Re_c}\right \vert^{-1}
\label{eq:MFanalog}
\end{equation}
with different prefactors below and above $Re_c$. This expression
can be interpreted as the combination of the Castaing statistical
temperature of turbulence $T \sim 1/\log Re$ with the classical
mean-field-theory expression for susceptibility at a critical
point:
\begin{equation*}
\tilde{\chi} \propto \left \vert T-T_c \right \vert^{-1},
\label{eq:MFpred}
\end{equation*}
Expression (\ref{eq:MFanalog}) has already been successfully
used in reference \cite{cortet2010} to describe the ``mean''
susceptibility $\chi$ computed with mean values ---instead of most
probable values for $\tilde{\chi}$--- of $I(t)$ and with a more
limited set of data. As already underlined, this rather fair
agreement between mean-field-theory and our turbulent measurements
is at first surprising. However, recent theoretical developments
suggest that in turbulent von K\'arm\'an flow the interactions are
long-range like \cite{naso1,naso2} supporting therefore the idea
that mean field theory is nearly applicable in our system.

\subsection{Momentization vs. magnetization}

Within the proposed analogy between our observations and the
para-ferromagnetism transition, the equivalence between the
magnetization and the momentization is the most questioning
aspect. In particular, the shape of the
$\overline{I}(\theta)$-curves is more similar to the magnetization
response curves of a single magnetic domains than to poly-domains.

For the para-ferromagnetism transition, the magnetization
increases with the distance below the critical temperature while
here, as stated in section \ref{stat}, the spontaneous
momentization is finite, limited in amplitude by $I_0$, and
decreases with the distance to $Re_c$. The system may share some
characteristics with reentrant noise-induced phase transition
similar to that observed in the annealed Ising model
\cite{thorpe76,genovese98}.  Our system is definitively of finite
size and it is thus coherent that both the level of fluctuation
$\sigma_I(Re_c)$ and the band of spontaneous fluctuations for
$I(t)$ ---which corresponds to a finite excursion of the shear
layer position along $z$--- remain finite at critical point
(figure \ref{zstag_reynolds}(c)).

In their von K\'{a}rm\'{a}n experiment at $\theta=0$ and $Re
\simeq 4 \times 10^5$, de la Torre and Burguete
(\cite{Torre2007,Burguete2009}) also observe  multistability of
three and only three states with can be interpreted, within our
framework, as $\overline{I}^\tau = -I_*$, $0$ and $+I_*$. While
the amplitude of the $\overline{I}^\tau = 0$ mode is constant
along time, their system switches between the two
$\overline{I}^\tau = \pm I_*$ modes with a very slow dynamics
(hours) and very fast  transitions. This dynamics does not seem to
depend on $Re$ once $Re \gtrsim 2  \times 10^4$, the symmetric
$\overline{I}^\tau = 0$ state being dominant below. Within our
framework, we could propose that, in this experiment, the
susceptibility is infinite whatever $Re$ above $2  \times 10^4$
(\emph{cf.} figure 5(a) of reference~\cite{Torre2007}).

\subsection{Specificity of turbulent fluctuations with respect to classical thermal noise}
\label{instabANDnoise}

In our turbulent system, the observed time-dynamics seems to be
due to the very high level of intrinsic turbulent fluctuations,
which are very different from classical thermal fluctuations that
pilot the dynamics of magnetic momentums. These turbulent
fluctuations are considered as the noise or the statistical
temperature origin of our system. In classical
close-to-equilibrium thermodynamical systems, thermal fluctuation
are generally considered as regular additive noise exactly as for
the classical treatment of instabilities in the presence of noise
where an external noise ---additive or multiplicative--- is
introduced as a perturbation of a regular system
\cite{review_noise}. For turbulent flows in general and for our
von K\'arm\'an flow in particular, the situation is different: the
flow itself is intrinsically highly fluctuating so that
fluctuations are of the same order of magnitude than the mean
values \cite{cortet09} and cannot be treated as perturbations. So,
the specific problem our experiment addresses is the behavior of
the instability of an average global quantity such as $I(t)$ in an
intrinsically fluctuating or noisy system. This is similar to the
problem of the experimental dynamo instability
\cite{dynamoP1,dynamoP5} which can only occur in turbulent flows
because of the nature of the existing conducting fluids of the
Universe. Therefore, non-classical behaviors are likely to be
observed for bifurcations or phase-transitions in such
highly-turbulent flows and our system represents a unique tool to
study these transitions.

\section*{Acknowledgments}

We thank S\'everine Atis and Lise Divaret for their participation
to the data acquisition. We thank C\'ecile Wiertel-Gasquet for the
interfacing and control of the experiment. We thank Kirone
Mallick, S\'ebastien Auma{\^i}tre, Javier Burguete and
Fr\'ed\'eric Moisy for fruitful discussions. PPC was supported by
Triangle de la Physique and EH by ANR SHREK (ANR-09-BLAN-0094-03).

\end{document}